\documentclass[preprint]{ptptex}

\usepackage{here}

\title{
Modification of the effective potential of $\phi^4$ theory using the Boltzmann factor
}

\author{
Kohzo \textsc{Nishida}%
\footnote{E-mail: EZF01671@nifty.com} 
}

\inst{
Department of Physics, Kyoto Sangyo University, Kyoto 603-8555, Japan 
}

\abst{
The need for a cutoff in the Lamb shift calculation suggests that high-energy virtual photons do not interact with the real particles.
In this study, we assume that  the creation of high-energy virtual particles is suppressed by a Boltzmann factor $e^{-\beta E}$
and apply this Boltzmann factor to the effective potential of the $\phi^4$ theory.
Consequently, the effective potential is transformed into 
$V(\phi) = 1/4!\phi^4 + \beta^{-4} \left \{ \phi^3 K_3(\phi) -\phi^2 K_2 (\phi) \right\}$.
$K_n(x)$ is the modified Bessel function of the second kind of order $n$. 
We demonstrate that the minimum of the effective potential occurs for $\phi \ne 0$.
}

\usepackage[]{graphicx}

\usepackage{caption}

\begin{document}

\maketitle

\section{Introduction}

In quantum field theory, cutoffs are often introduced when calculating physical quantities.
Let us demonstrate this in a Lamb shift\cite{1,2,3,4,5} calculation.
The Lamb shift $\Delta E$ between $2s_{1/2}$ and $2p_{1/2}$ levels is denoted\cite{6}
\begin{equation}
\label{eq:1}
\Delta E = \frac{\alpha^5 m_e}{6\pi} \int^{m_e}_{1/a_0} \frac{d p}{p}
= \frac{\alpha^5 m_e}{6\pi}  \ln \frac{1}{\alpha},
\end{equation}
where $m_e$ is the mass of the electron, $\alpha$ is the fine structure constant, and $a_0=1/(m_e\alpha)$ is the Bohr radius.
The energy integral of the virtual photon, $\int d p/p$, stops counting photons when their wavelengths surpass the size of the atom, $a_0$.  
On the short-wavelength side, this integral stops counting photons when their wavelengths are less than the Compton wavelength. 
That is, a cutoff $m_e$ is introduced in the Lamb shift calculation.


Renormalization by cutoff is also an operation that does not count virtual particles with energies larger than the cutoff.
Thus, the agreement between the experimental and theoretical values obtained by introducing the cutoff suggests that 
high-energy virtual photons do not affect real particles virtually.

It is known that Eq. (\ref{eq:1}) with a cutoff $m_e$ can be approximated as an integral with an infinite range using a smooth cutoff function $e^{-p/m_e}$, 
which is stated as follows:
\begin{equation}
\label{eq:2}
\Delta E \simeq \frac{\alpha^5 m_e}{6\pi} \int^{\infty}_{1/a_0} \frac{d p}{p} e^{-p/m_e}
\simeq  \frac{\alpha^5 m_e}{6\pi}  \left ( \ln \frac{1}{ \alpha} -\gamma + \pi \alpha \right ),
\end{equation}
where the following integral formula was used:
\begin{equation}
\label{eq:3}
\int^{\infty}_{x} \frac{e^{-p} }{p}d p = \Gamma(0, x) = (-\log(x) -\gamma) + x-\frac{x^2}{4} + \cdots,
\end{equation}
where $\Gamma(s, x)$ denotes the upper incomplete gamma function, and 
$\gamma=0.57721\cdots$ is the Euler constant.
Thus, we can approximate an integral with a cutoff $\Lambda$ as an integral with an infinite integral range multiplied by the smooth cutoff function $e^{-p/\Lambda}$.

The cutoff function $e^{-p/\Lambda}$ takes the same form as the Boltzmann factor $e^{-\beta E}$.
If the cutoff function is a Boltzmann factor, 
the creation of virtual particles follows a Boltzmann distribution.
In our previous  research\cite{7,8}, we investigated how the Boltzmann factor affects the quantum field theory,
and obtained a finite zero-point energy, modified Coulomb potential, etc.
In this study, we apply the Boltzmann factor to the effective potential of the $\phi^4$ theory 
and demonstrate that the minimum of the effective potential occurs for $\phi \ne 0$.

\section{Modification of the effective potential}

We consider the massless $\phi^4$ theory with the Lagrangian
\begin{equation}
\label{eq:4}
{\cal L} = \frac{1}{2}\partial_\mu \phi(x) \partial^\mu \phi(x) -\frac{\lambda}{4!}\phi^4
\end{equation}
with $\lambda > 0$.
It is known that the no-loop and one-loop effective potentials for Eq. (\ref{eq:4}) are given by
\begin{equation}
\label{eq:5}
V(\phi_c) = \frac{\lambda}{4!}\phi_c^4 + V_1(\phi_c),
\end{equation}
where
\begin{equation}
\label{eq:6}
V_1(\phi_c) 
=
\frac{1}{2} \int_{-\infty}^\infty \frac{d^4p}{i(2\pi)^4} \ln \left(\frac{\lambda}{2}\phi_c^2-p^2-i\epsilon \right).
\end{equation}
Here, $\phi_c$  denotes the vacuum expectation value of $\phi$.
The integral in Eq. (\ref{eq:6}) includes the contribution of high-energy
virtual particles and therefore diverges.
If the creation of high-energy virtual particles is suppressed by the Boltzmann factor, Eq. (\ref{eq:6}) should be modified as follows:

\begin{equation}
\label{eq:7}
V_1(\phi_c) 
=
 \frac{1}{2} \int_{-\infty}^\infty \frac{d^4p}{i(2\pi)^4} \ln \left(\frac{\lambda}{2}\phi_c^2-p^2-i\epsilon \right)
\times
 e^{-\beta E(\boldsymbol{p}) } ,
\,\,\,\,\,\,
 E(\boldsymbol{p}) = \sqrt{\boldsymbol{p}^2+ \frac{\lambda}{2} \phi_c^2},
\end{equation}
where $\beta$ is a constant with a mass dimension of $-1$, and we interpret it as the reverse temperature. 
In regular quantum field theory, $\beta^{-1}$ is the cutoff parameter. 
In Section 3, we give the detailed derivation of  Eq. (\ref{eq:7}).

Performing the $p^0$ integral, we have  (see Appendix A for details)
\begin{equation}
\label{eq:8}
V_1(\phi_c) =  
\int_{-\infty}^\infty \frac{d^3p} {(2\pi)^3} \frac{1}{2} \sqrt{\boldsymbol{p}^2 + \frac{\lambda}{2} \phi_c^2} 
\times
 e^{-\beta E(\boldsymbol{p}) } ,
\end{equation}
Let us calculate the integral in Eq. (\ref{eq:8}).
Using the integral formula (see Appendix B for details)
\begin{equation}
\label{eq:9}
\int_x^\infty y^2 \sqrt{ y^2 -x^2} e^{-y} dy.
=
x^3 K_3(x) -x^2 K_2(x),
\end{equation}
we have 
\begin{eqnarray}
\label{eq:10}
V_1(\phi_c) 
&=&
\int_{-\infty}^\infty \frac{d^3p} {(2\pi)^3} \frac{1}{2} \sqrt{\boldsymbol{p}^2 + \frac{\lambda}{2} \phi_c^2} 
\times
 e^{-\beta \sqrt{\boldsymbol{p}^2 + \lambda/2 \phi_c^2  } } \nonumber \\
&=& 
\frac{\beta^{-4}  }{(2\pi)^2} \int_{x}^{\infty}   y^2   \sqrt{y^2 -x^2 } e^{-y }  dy  
\,\,\,\,\,\,(y^2=\beta^2 \boldsymbol{p}^2+x^2, \,\, x=\beta \sqrt{\frac{\lambda}{2}}\phi_c) \nonumber \\
&=& 
\frac{\beta^{-4}   }{(2\pi)^2} \left \{ x^3 K_3(x) -x^2 K_2(x) \right\}, 
\end{eqnarray}
where $K_n(x)$ is the modified Bessel function of the second kind of order $n$.
Finally, we obtain the modified effective potential as follows:
\begin{eqnarray}
\label{eq:11}
 V(\phi_c) 
 &=&
 \frac{\lambda}{4!}\phi_c^4
\nonumber \\
 &+&
\frac{\beta^{-4}   }{(2\pi)^2} \left \{ 
\left( \beta \sqrt{\frac{\lambda}{2}} |\phi_c| \right)^3 K_3\left( \beta \sqrt{\frac{ \lambda}{2}} |\phi_c| \right) 
-\left(\beta \sqrt{\frac{\lambda}{2}} |\phi_c| \right)^2 K_2\left(\beta \sqrt{\frac{\lambda}{2}} |\phi_c| \right) \right\},
\end{eqnarray}
where we take the absolute value of $\phi_c$ because Eq. (\ref{eq:8}) is invariant under $\phi_c \rightarrow -\phi_c$.

If we replace the variable $\phi_c$ with $x=\beta \sqrt{\frac{\lambda}{2}}\phi_c$, 
we have
\begin{equation}
\label{eq:12}
V(x) = \beta^{-4} \left [\frac{1}{3! \lambda} x^4 + \frac{1 }{(2\pi)^2} \left \{ |x|^3 K_3(|x|) -|x|^2 K_2 (|x|) \right\} \right].
\end{equation}
We present the graph of Eq. (\ref{eq:12}) in Fig. 1, where we set $\beta=1$ and $\lambda=1$.
From Fig. 1, we can confirm that 
the minimum of the effective potential occurs for $\phi_c \ne 0$.
Therefore, the vacuum expectation value of $\phi$ is not zero because of the spontaneous symmetry breaking.

\begin{figure}[H]
\label{my_fig1}
\begin{center}
\includegraphics[width=120mm]{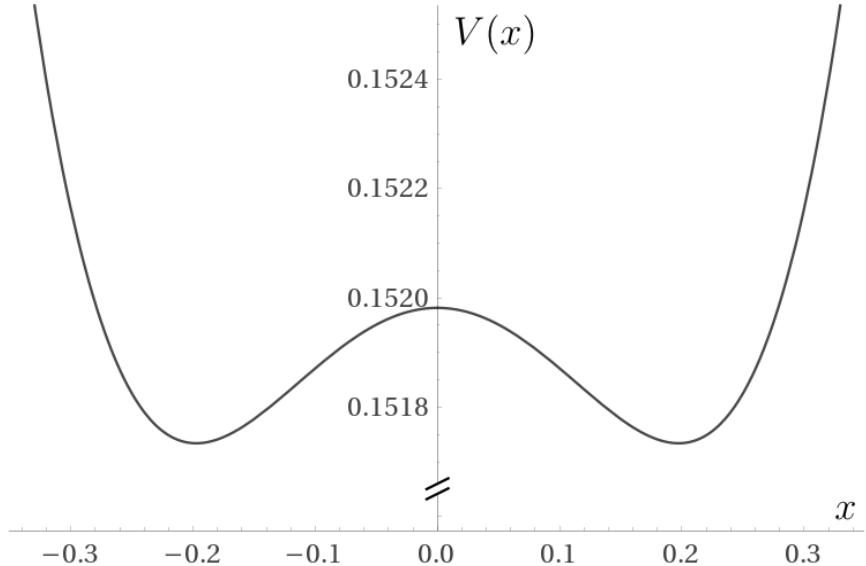}
\caption{
The modified effective potential 
$V(x) = \beta^{-4} \left [\frac{1}{3! \lambda}x^4 + \frac{1 }{(2\pi)^2} \left \{ |x|^3 K_3(|x|) -|x|^2 K_2 (|x|) \right\} \right]$
with $\beta=1$ and $\lambda=1$.
$K_n(x)$ is the modified Bessel function of the second kind of order $n$.
 }
\end{center}
\end{figure}

Finally, let us examine the extremum of $ V(x)$ for $x \geq 0$ obtained in Eq. (\ref{eq:12}).
Using the derivative formula of the modified Bessel function of the second kind:
\begin{equation}
\label{eq:13}
\frac{d}{dx} (x^n K_n(x)) = -x^n K_{n-1}(x),
\end{equation}
 we obtain 
\begin{equation}
\label{eq:14}
\frac{d}{dx} V(x) =  \frac{\beta^{-4} }{ 4\pi^2 } x^2
 \left\{ \frac{8\pi^2}{3\lambda} x -x K_2(x) + K_1(x) \right \}.
\end{equation}
Therefore, $V(x)$ takes a local minimum at $x$ that satisfies the following equation:
\begin{equation}
\label{eq:15}
\frac{8\pi^2}{3\lambda} x -x K_2(x) + K_1(x) = 0.
\end{equation}
The local maximum of $V(x)$ is at $x=0$, giving
\begin{equation}
\label{eq:16}
V(0)=V_1(0) =
\int_{-\infty}^\infty \frac{d^3p} {(2\pi)^3} \frac{1}{2} \sqrt{\boldsymbol{p}^2} 
\times
e^{-\beta \sqrt{\boldsymbol{p}^2} }
=
\frac{3\beta^{-4} }{2\pi^2}.
\end{equation}

\section{Modification of the path integral quantization}

Our objective in this section is to obtain Equation (\ref{eq:7}) by modified path integral quantization.
To illustrate our idea, let us consider the momentum representation of the path integral:
\begin{equation}
\label{eq:17}
\int \mathcal{D} \phi(p) \exp i \int d^4 p \left[  {\cal L}(\phi(p))+J(p) \phi(p) \right]
=
\prod_p^\infty \left\{
 \int d \phi(p) \exp i \Delta^4 p \left[  {\cal L}(\phi(p))+J(p) \phi(p) \right]
\right \}.
\end{equation}
$\prod_p^\infty$ in Eq. (\ref{eq:17}) includes the contribution of high-energy
virtual particles.
If the creation of high-energy virtual particles is suppressed by the Boltzmann factor, Eq. (\ref{eq:17}) should be modified as follows:
\begin{eqnarray}
\label{eq:18}
\lefteqn{
\prod_p^\infty \left\{
 \int d \phi(p) \exp i \Delta^4 p \left[  {\cal L}(\phi(p))+J(p) \phi(p) \right]
\right \}
}
\nonumber \\
&\Rightarrow&
\prod_p^\infty \left\{
 \int d \phi(p) \exp i \Delta^4 p \left[  {\cal L}(\phi(p))+J(p) \phi(p) \right]
\right \}^{\exp(-\beta|\bar{u}_\mu p^\mu(\boldsymbol{p})|)}
\nonumber \\
&\simeq&
\prod_p^{\beta^{-1}} \left\{
 \int d \phi(p) \exp i \Delta^4 p \left[  {\cal L}(\phi(p))+J(p) \phi(p) \right]
\right \},
\end{eqnarray}
where $\exp(-\beta|\bar{u}_\mu p^\mu(\boldsymbol{p})|)$ is a relativistic Boltzmann factor and 
$p^\mu(\boldsymbol{p}) = \left( E(\boldsymbol{p}),\boldsymbol{p} \right)$. 
We assume that $E(\boldsymbol{p})$ satisfies the mass shell condition.
$\bar{u}_\mu$ is an  average of four-velocities of particles created by the scalar field $\phi(x)$.
The spatial distribution of matter in the universe is homogeneous and isotropic, 
which means that the average  created  particle four-velocities $\bar{u}^\mu$ can be written in any coordinate system as 
\begin{equation}
\label{eq:19}
\bar{u}_\mu = (1,0,0,0),
\end{equation}
because the spatial component is canceled with plus and minus appearing equally.
So in any coordinate system, we can always rewrite the relativistic Boltzmann factor to
\begin{equation}
\label{eq:20}
e^{-\beta|\bar{u}_\mu p^\mu(\boldsymbol{p})| }
=
e^{-\beta E(\boldsymbol{p}) }.
\end{equation}
Hereinafter, 
we describe the relativistic Boltzmann factor as $e^{-\beta E(\boldsymbol{p}) }$.

Let us apply the idea of Eq. (\ref{eq:18}) to the scalar field theory.
First, we consider the free-field generating functional of Green functions:
\begin{equation}
\label{eq:21}
Z_{0}[J]=N\int \mathcal{D} \phi(x) \exp i \int d^4 x \left[ -\frac{1}{2} \phi(x) (\Box + \mu^2)\phi(x) +J(x) \phi(x) \right].
\end{equation}
The Fourier expansions of $\phi(x)$ and $J(x)$ by momentum-energy are 
\begin{equation}
\label{eq:22}
\phi(x) =  \frac{1}{ \sqrt{(2\pi)^4} } \int  \frac{1}{ \sqrt{2}} \{ a(p) e^{-ip x} + a^*(p) e^{ip x} \} d^4 p
\end{equation}
and
\begin{equation}
\label{eq:23}
J(x) = \frac{1}{ \sqrt{(2\pi)^4} } \int \frac{1}{\sqrt{2}} \{ \hat{J} (p) e^{-ip x} +\hat{J}^* (p) e^{ip x} \} d^4 p,
\end{equation}
respectively.
By inserting Eq. (\ref{eq:22}) and Eq. (\ref{eq:23}) into Eq. (\ref{eq:21}), 
Eq. (\ref{eq:21}) can be written  in momentum-energy space as 
\begin{eqnarray}
\label{eq:24}
\lefteqn{ Z_{0}[J] =}
\nonumber \\
&=&
 \left \{
\int \mathcal{D} a^*(p) \mathcal{D} a(p)  \exp i 2\int d^4 p \left[ -a^*(p) (\mu^2-p^2) a(p) +\hat{J}^* (p) a(p) +\hat{J}(p) a^*(p) \right]
\right \}^{1/2}
\nonumber \\
&=&
\prod_p^\infty \left\{ \int d a^*(p) d a(p)  \exp i 2\Delta^4 p \left[ -a^*(p) (\mu^2-p^2) a(p) +\hat{J}^* (p) a(p) +\hat{J}(p) a^*(p) \right] \right\}^{1/2},
\nonumber \\
\end{eqnarray}
where we used $a(-p)=a^*(p), a^*(-p)=a(p)$.
According to our idea of Eq. (\ref{eq:18}), Eq. (\ref{eq:24}) should be modified as follows:
\begin{eqnarray}
\label{eq:25}
\lefteqn{ \text{(\ref{eq:24})} \Rightarrow Z_{0}[J]  =}
\nonumber \\
&&
\prod_p^\infty \left\{ \int d a^*(p) d a(p)  \exp i 2 \Delta^4 p\left[ -a^*(p) (\mu^2-p^2) a(p) +\hat{J}^* (p) a(p) +\hat{J}(p) a^*(p) \right] \right\}^{1/2\exp(-\beta E(\boldsymbol{p}))}.
\nonumber \\
\end{eqnarray}
Here we assume that $E(\boldsymbol{p})$ satisfies the mass shell condition 
$E^2(\boldsymbol{p})=\boldsymbol{p}^2+\mu^2$ 
derived from the equation of motion of the free scalar field $\phi$.
By the change of variables $a'=a-iJ/(iM)$, Eq. (\ref{eq:25}) becomes
\begin{eqnarray}
\label{eq:26}
\lefteqn{ Z_{0}[J] =}
\nonumber \\
&&
\prod_p^\infty \left\{ \int d a'^*(p) d a'(p)  \exp 2\Delta^4 p \left[ -a'^*(p) iM a'(p) - (i\hat{J})^*(p) (iM)^{-1} i\hat{J}(p)  \right] \right\}^{1/2\exp(-\beta E(\boldsymbol{p}))}
\end{eqnarray}
with $M=\mu^2-p^2$.
Using the Gauss integral formula
\begin{equation}
\label{eq:27}
\int dz^* dz \exp[-Az^*z]  = \frac{2\pi}{A},
\end{equation}
we have
\begin{eqnarray}
\label{eq:28}
Z_{0}[J]
&=&
N \prod_p \left\{   \frac{2\pi}{ i(\mu^2-p^2) 2 \Delta^4 p}  
\exp \left[  - 2\Delta^4 p (i\hat{J})^*(p) \frac{ 1}{ i(\mu^2-p^2)} i\hat{J}(p) \right] \right\}^{1/2\exp(-\beta E(\boldsymbol{p}))}
\nonumber \\
&=&
Z_{0}[J = 0]
 \exp \left[ -i \int d^4 p  \hat{J}^* (p) \frac{ e^{-\beta E(\boldsymbol{p})} }{ p^2 -\mu^2} \hat{J}(p) \right].
\end{eqnarray}
By the reverse Fourier transform
\begin{equation}
\label{eq:29}
{\hat J}(p) = \frac{1} { \sqrt{(2\pi)^4}} \int \frac{1}{\sqrt{2}} J(x) e^{ipx} d^4 x,
\end{equation}
we get
\begin{equation}
\label{eq:30}
Z_{0}[J]
=
Z_{0}[J = 0]
\exp \left[ \frac{-i}{2}\int d^4 x \int d^4 y J(x) 
\int \frac{d^4 p}{  (2\pi)^4} \frac{ e^{-\beta E(\boldsymbol{p})}  }{ p^2 -\mu^2 + i\epsilon} e^{ -ip(x-y)} 
J(y) \right], 
\end{equation}
where we used the epsilon prescription, $\mu^2 \rightarrow \mu^2 -i\epsilon$.
Thus, we obtain the modified Feynman propagator
\begin{eqnarray}
\label{eq:31}
i\Delta_{F}(x-y) 
&=&
\left.
 -\frac{ 1}{Z_{0}[J = 0]} 
\frac{ \delta^2 Z_{0}[J] }{ \delta J(x) \delta J(y)} 
\right |_{J=0}
\nonumber \\
&=&
\int \frac{d^4 p}{  (4\pi)^2} \frac{ e^{-\beta E(\boldsymbol{p})}  }{ p^2 -\mu^2 + i\epsilon} e^{ -ip(x-y)} .
\end{eqnarray}
 Eq. (\ref{eq:31}) is in agreement with the Feynman propagator obtained by the modified canonical quantization in Ref. [7].

Next, we consider the generating functional for interacting fields:
\begin{equation}
\label{eq:32}
Z[J]
=
\int \mathcal{D} \phi(x) \exp i \int d^4 x\left[ {\cal L}( \phi(x) )+J(x) \phi(x) \right]
\end{equation}
with the Lagrangian
\begin{equation}
\label{eq:33}
{\cal L}(\phi(x)) = -\frac{1}{2} \phi(x) (\Box + \mu^2)\phi(x) -\frac{\lambda}{4!}\phi^4(x).
\end{equation}
The generating functional of connected Green functions, $W[J]$, is given by
\begin{equation}
\label{eq:34}
iW[J]
=
\ln Z[J]. 
\end{equation}
The effective action $\Gamma[\phi_c]$ is defined by the functional Legendre transform:
\begin{equation}
\label{eq:35}
\Gamma[\phi_c]
=
W[J] - \int d^4 x J(x) \phi_c(x),
\end{equation}
where
$\phi_c(x)$ is the vacuum expectation value of $\phi(x)$ in the presence of the source $J(x)$:
\begin{equation}
\label{eq:36}
\phi_c(x) = \frac{\delta}{ \delta J(x)} W[J]
=
\left[ \frac{<0|\phi(x)|0>}{<0|0>} \right]_J.
\end{equation}
Therefor, we find
\begin{eqnarray}
\label{eq:37}
\Gamma[\phi_c] 
&=&
W[J] - \int d^4 x J(x) \phi_c(x)
\nonumber \\
&=&
-i\ln \int \mathcal{D} \phi(x) \exp i \int d^4 x\left[ {\cal L}(x)+J(x) (\phi(x)-\phi_c(x)) \right].
\end{eqnarray}
If we make the substitution 
\begin{equation}
\label{eq:38}
\phi \rightarrow \phi + \phi_c,
\end{equation}
we have
\begin{eqnarray}
\label{eq:39}
\Gamma[\phi_c] 
&=&
-i\ln \int \mathcal{D} \phi(x) \exp i \int d^4 x\left[ {\cal L}(\phi + \phi_c)+J(x) \phi_c(x) \right].
\end{eqnarray}
By expanding ${\cal L}(\phi + \phi_c)$  around $\phi_c$
in terms of $\phi$ to the one-loop order,
we obtain
\begin{equation}
\label{eq:40}
\Gamma[\phi_c]
=
\int d^4 x {\cal L}(\phi_c) + \Gamma_1 [\phi_c],
\end{equation}
with
\begin{equation}
\label{eq:41}
\Gamma_1 [\phi_c]
=
-i \ln \int \mathcal{D} \phi(x) \exp i \int d^4 x \left[ -\frac{1}{2}\phi(x)(\Box + \mu^2 +\frac{\lambda}{2}\phi_c^2(x)) \phi(x) \right] 
\end{equation}

Hereinafter, we consider the case where $\phi_c$ is a spacetime-independent constant.
By the Fourier transform,
\begin{equation}
\label{eq:42}
\phi(x) =  \frac{1}{ \sqrt{(2\pi)^4} } \int  \frac{1}{ \sqrt{2}} \{ a(p) e^{-ip x} + a^*(p) e^{ip x} \} d^4 p,
\end{equation}
Eq. (\ref{eq:41})  is calculated as
\begin{equation}
\label{eq:43}
\Gamma_1 [\phi_c]
=
-i \ln
\prod_p^\infty \left\{ 
\int d a^*(p) d a(p)  \exp i 2\left[ -a^*(p) (\mu^2+\frac{\lambda}{2}\phi_c^2-p^2) a(p)  \right] 
\right\}^{1/2}.
\end{equation}
According to our idea of Eq. (\ref{eq:18}), Eq. (\ref{eq:43}) should be modified as follows:
\begin{eqnarray}
\label{eq:44}
\lefteqn{ \text{(\ref{eq:43})} \Rightarrow \Gamma_1 [\phi_c]}
\nonumber \\
&=&
-i \ln
\prod_p^\infty \left\{ 
\int d a^*(p) d a(p)  \exp i 2\left[ -a^*(p) (\mu^2+\frac{\lambda}{2}\phi_c^2-p^2) a(p)  \right] 
\right\}^{1/2 \exp(-\beta E(\boldsymbol{p}))}.
\end{eqnarray}
Here we assume that $E(\boldsymbol{p})$ satisfies the mass shell condition 
$E^2(\boldsymbol{p})= \boldsymbol{p}^2+\mu^2+\lambda/2\phi_c^2 $
derived from the equation of motion of the free scalar field $\phi$.
Notice that $\phi_c$ change the effective mass of $\phi$.
Using the Gauss integral formula Eq. (\ref{eq:27}), 
we can calculate Eq. (\ref{eq:44}) as follows:
\begin{eqnarray}
\label{eq:45}
\lefteqn{ \Gamma_1 [\phi_c]}
\nonumber \\
&=&
-i \ln
\prod_p^\infty \left\{ 
\frac{2\pi }{i(\mu^2 +\lambda/2 \phi_c^2-p^2) 2\Delta^4 p }
\right\}^{1/2\exp(-\beta E(\boldsymbol{p}))}
\nonumber \\
&=&
-i \frac{1}{\Delta^4 p} \frac{1}{2} \sum_p^\infty \Delta^4 p  e^{-\beta E(\boldsymbol{p})} \ln
\left\{ 
\frac{2\pi }{i(\mu^2 +\lambda/2 \phi_c^2-p^2) 2\Delta^4 p}
\right\}
\nonumber \\
&=&
\frac{i}{2} (2\pi)^4 \delta^4 (0) \int \frac{d^4 p}{(2\pi)^4}  e^{-\beta E(\boldsymbol{p})} \left \{
\ln \left( 
\mu^2 +\frac{\lambda}{2} \phi_c^2-p^2
\right)
-\ln \left( \frac{\pi}{i\Delta^4 p} \right)
\right \}
\nonumber \\
&=&
 \left \{
\frac{i}{2} \int \frac{d^4 p}{(2\pi)^4}  e^{-\beta E(\boldsymbol{p})} 
\ln \left( 
\mu^2 +\frac{\lambda}{2} \phi_c^2-p^2
\right)
+\text{constant}
\right \}
\int d^4 x,
\end{eqnarray}
where we used $(2\pi)^4/\Delta^4 p = (2\pi)^4 \delta^4(0) = \left. \int d^4 x e^{ikx} \right|_{k=0} = \int d^4x$.
The effective potential $V(\phi_c)$ is defined as 
\begin{equation}
\label{eq:46}
\Gamma[\phi_c] = -V(\phi_c)\int d^4 x.
\end{equation}
Therefore, we finally obtain the modified effective potential:
\begin{equation}
\label{eq:47}
V(\phi_c) = 
\frac{1}{2}\mu^2\phi_c^2 + \frac{\lambda}{4!} \phi_c^4
+
V_1(\phi_c)
\end{equation}
with 
\begin{equation}
\label{eq:48}
V_1(\phi_c) = 
\frac{1}{2}\int^\infty_{\infty} \frac{d^4 p}{ i(2\pi)^4}  e^{-\beta E(\boldsymbol{p})} 
\ln \left( 
\mu^2 +\frac{\lambda}{2} \phi_c^2-p^2 - i\epsilon
\right),
\,\,\,\,\,\,
 E(\boldsymbol{p}) = \sqrt{\boldsymbol{p}^2+ \mu^2 + \frac{\lambda}{2} \phi_c^2},
\end{equation}
where we used the epsilon prescription, $\mu^2 \rightarrow \mu^2 -i\epsilon$, 
and ignored the constant term.
If we take $\mu=0$, the aim of this section to derive Eq. (\ref{eq:7}) has now  been achieved.

\appendix

\section{Non-relativistic expression for the effective potential}

In this Appendix we derive  Eq. (\ref{eq:8}) from Eq. (\ref{eq:7}).
%
We can calculate  Eq. (\ref{eq:7}) by transforming into the following:

\begin{eqnarray}
\label{eq:A1}
\lefteqn{ V_1(\phi_c)  }
\nonumber \\
&=&
\frac{1}{2} \int_{-\infty}^\infty \frac{d^4p}{i(2\pi)^4} 
e^{-\beta E(\boldsymbol{p})}
\ln \left(\frac{\lambda}{2}\phi_c^2-p^2-i\epsilon \right)
\nonumber \\
&=&
\left. 
\frac{1}{2}\int \frac{d^4 p}{ i(2\pi)^4} 
e^{-\beta E(\boldsymbol{p})} 
\ln \left( 
x-p^2-i\epsilon
\right)
\right |_{ x = \frac{\lambda}{2}\phi_c^2}
\nonumber \\
&=&
\left. 
 \frac{1}{2} \int_{-\infty}^\infty \frac{d^4p} {i(2\pi)^4} 
e^{ -\beta E(\boldsymbol{p})}
\int
 dx  \frac{1 }{x -p^2 -i\epsilon}
\right |_{ x = \frac{\lambda}{2}\phi_c^2}
\nonumber \\
&=&
\left.  
\frac{1}{2} \int_{-\infty}^\infty \frac{d^3p} {i(2\pi)^4} 
e^{ -\beta E(\boldsymbol{p})}
\int dx
\int_C d p^0   \frac{-1 } { (p^0- \sqrt{x +\boldsymbol{p}^2 } + i\epsilon)(p^0+\sqrt{x +\boldsymbol{p}^2 } - i\epsilon) } 
\right |_{ x = \frac{\lambda}{2}\phi_c^2}
\nonumber \\
&=&
\left.
\frac{1}{2} \int_{-\infty}^\infty \frac{d^3p} {i(2\pi)^4} 
e^{ -\beta E(\boldsymbol{p})}
\int dx
\frac{ -2\pi i  } { p^0-  \sqrt{x +\boldsymbol{p}^2 }+ i\epsilon} 
\right|_{p^0 = -\sqrt{ x +\boldsymbol{p}^2 }+i\epsilon, x = \frac{\lambda}{2}\phi_c^2}
\nonumber \\
&=&
\left.  
\frac{1}{2} \int_{-\infty}^\infty \frac{d^3p} {(2\pi)^3} 
e^{ -\beta E(\boldsymbol{p})}
\int dx
\frac{  1 }{2\sqrt{x +\boldsymbol{p}^2 }}
\right |_{ x = \frac{\lambda}{2}\phi_c^2}
\nonumber \\
&=&
\int_{-\infty}^\infty \frac{d^3p} {(2\pi)^3} 
\frac{1}{2} 
\sqrt{ \boldsymbol{p}^2 + \frac{\lambda}{2}\phi_c^2 }
\times
e^{ -\beta E(\boldsymbol{p})}
\end{eqnarray}
Thus, we can obtain Eq. (\ref{eq:8}).

\section{Integral formula}
In this Appendix we derive  Eq. (\ref{eq:9}).
We use the following identity of the modified Bessel function of the second kind $K_n(x)$ of order $n$:
\begin{equation}
\label{eq:B1}
K_n(x) = 
\frac{\sqrt{\pi}} { (n-1/2)! } \left( \frac{1}{2} x \right)^n \int_1^\infty e^{-xt} (t^2 -1)^{n-1/2} dt
\,\,\,\,\,\,
\mbox{for}
\,\,
n > -\frac{1}{2}.
\end{equation}
If we set $xt=y$,  
\begin{equation}
\label{eq:B2}
K_n(x) 
=
\frac{\sqrt{\pi}} { (n-1/2)! } \frac{1}{2^n} \frac{1}{x^n} \int_x^\infty e^{-y} (y^2 -x^2)^{n-1/2} dy.
\end{equation}
Therefore, 
\begin{equation}
\label{eq:B3}
\int_x^\infty  (y^2 -x^2)^{n-1/2} e^{-y} dy
=
\frac{2^n(n-1/2)!}{\sqrt{\pi}} x^n K_n(x).
\end{equation}
Integration by parts yields
\begin{equation}
\label{eq:B4}
\int_x^\infty y(y^2 -x^2)^{n-3/2} e^{-y} dy
=
\frac{2^{n-1}(n-3/2)!}{\sqrt{\pi}} x^n K_n(x).
\end{equation}
%
Furthermore, the integration by parts in the left-hand side of Eq. (\ref{eq:B4}) gives
\begin{eqnarray}
\label{eq:B5}
\lefteqn{ \int_x^\infty y(y^2 -x^2)^{n-3/2} e^{-y} dy }
\nonumber \\
&=&
\frac{2^{n-1}(n-3/2)!}{\sqrt{\pi}} x^{n-1} K_{n-1}(x)
+2(n-\frac{3}{2}) \int_x^\infty y^2(y^2 -x^2)^{n-5/2} e^{-y} dy.
\end{eqnarray}
Therefore, 
\begin{eqnarray}
\label{eq:B6}
\lefteqn{ \int_x^\infty y^2(y^2 -x^2)^{n-5/2} e^{-y} dy}
\nonumber \\
&=&
\frac{2^{n-2}(n-5/2)!}{\sqrt{\pi}} x^{n} K_{n}(x)
-
\frac{2^{n-2}(n-5/2)!}{\sqrt{\pi}} x^{n-1} K_{n-1}(x).
\end{eqnarray}
At $n=3$, we finally obtain Eq. (\ref{eq:9}):
\begin{equation}
\label{eq:B7}
\int_x^\infty y^2 \sqrt{ y^2 -x^2} e^{-y} dy.
=
x^3 K_3(x) -x^2 K_2(x).
\end{equation}

\end{document}